\documentclass[acmsmall]{acmart}

\usepackage[utf8]{inputenc}
\usepackage{multirow}
\usepackage{tabularx}
\usepackage{booktabs} 
\usepackage{hyperref}
\usepackage{amsmath}
\usepackage{verbatim}
\usepackage{siunitx}
\usepackage[shortlabels]{enumitem}
\usepackage{fancybox}
\usepackage{url}
\usepackage{graphicx}
\usepackage{setspace}
\usepackage{soul}
\usepackage[roman]{parnotes}
\usepackage{color}
\usepackage{colortbl}
\usepackage{listings}
\usepackage{subcaption}
\usepackage{algorithm, algpseudocode}
\algnewcommand\algorithmicto{\textbf{to}}
\algnewcommand\RETURN{\State \textbf{return} }
\usepackage{textcomp}
\usepackage{xcolor}
\usepackage[justification=centering]{caption}
\usepackage{pifont}
\usepackage{soul}
\usepackage{multirow}
\newcounter{observation}
\newcommand{\observation}[1]{\refstepcounter{observation}
	\begin{center}
		\Ovalbox{
			\begin{minipage}{0.95\columnwidth}
				{\bf Observation:} #1
			\end{minipage}
		}
	\end{center}
}

\AtBeginDocument{%
  \providecommand\BibTeX{{%
    Bib\TeX}}}

\acmBooktitle{Proceedings of the 32nd ACM Symposium on the Foundations of Software Engineering (FSE '24), November 15--19, 2024, Porto de Galinhas, Brazil}
\acmPrice{15.00}
\acmISBN{978-1-4503-XXXX-X/18/06}

\begin{document}

\title{Does the Order of Fine-tuning Matter and Why?}

\author{Qihong Chen}
\email{chenqh@uci.edu}
\orcid{1234-5678-9012}
\author{Jiawei Li}
\authornotemark[1]
\email{jiawl28@uci.edu}
\affiliation{%
  \institution{University of California, Irvine}
  \city{Irvine}
  \state{CA}
  \country{USA}
}

\author{Hyunjae Suh}
\affiliation{%
  \institution{University of California, Irvine}
  \city{Irvine}
  \state{CA}
  \country{USA}
}
\email{hyunjas@uci.edu}

\author{Lianghao Jiang}
\affiliation{%
  \institution{University of California, Irvine}
  \city{Irvine}
  \state{CA}
  \country{USA}
}
\email{lianghaj@uci.edu}

\author{Zheng Zhou}
\affiliation{%
  \institution{University of California, Irvine}
  \city{Irvine}
  \state{CA}
  \country{USA}
}
\email{zhengz22@uci.edu}

\author{Jingze Chen}
\affiliation{%
  \institution{University of California, Irvine}
  \city{Irvine}
  \state{CA}
  \country{USA}
}
\email{jingzec@uci.edu}

\author{Jiri Gesi}
\affiliation{%
  \institution{Amazon}
  \city{Palo Alto}
  \state{CA}
  \country{USA}
}
\email{jirigesi@amazon.com}

\author{Iftekhar Ahmed}
\affiliation{%
  \institution{University of California, Irvine}
  \city{Irvine}
  \state{CA}
  \country{USA}
}
\email{iftekha@uci.edu}

\renewcommand{\shortauthors}{Chen et al.}


\begin{abstract}
To improve the performance on a target task, researchers have fine-tuned language models with an intermediate task before the target task of interest. However, previous works have focused on the pre-trained language models and downstream tasks in Natural Language Processing (NLP) and considered only one intermediate task. The effect of fine-tuning multiple intermediate tasks and their ordering on target task performance has not been fully explored in Software Engineering. In this study, we perform the first empirical study on analyzing the impact of task ordering on target task performance. Experimental results show that there is an impact of task ordering on target task performance by up to 6\% of performance gain and up to 4\% of performance loss. To explain such an impact, we consider a variety of potential factors, including the characteristics of dataset (syntactic similarity and semantic similarity analysis, dataset size), model (probing task and attention analysis), and task (task affinity analysis). Our study provides Software Engineering researchers and practitioners with insights into the effect of task orderings and how to select the one that is cost-effective while achieving the best performance gain. We provide our replication package in~\cite{replication}.

\end{abstract}




\maketitle


\section{Introduction}
\label{sec:intro}

\renewcommand*{\thefootnote}{\fnsymbol{footnote}}
\footnotetext[1]{The ﬁrst two authors contributed equally to this work.}
\renewcommand*{\thefootnote}{\arabic{footnote}}



The recent advancements in self-supervised learning, where annotated training data can be automatically generated from a text corpus, have increased the importance of large language models in automating various Natural Language Processing (NLP) tasks such as text classification and question answering. These models undergo pre-training on an extensive text corpus, during which their parameter weights are trained to encode general linguistic knowledge about the language~\cite{devlin2018bert}. To apply these models to downstream tasks, they undergo further fine-tuning on a task-specific dataset, which typically involves human-annotated training data that is expensive to acquire in a supervised manner. This entire process can be considered as a transfer learning approach where the linguistic knowledge acquired during the pre-training phase is transferred through the fine-tuning phase to the downstream task, resulting in an accelerated training process and improved model performance.


To further improve model performance on downstream tasks, researchers in NLP have investigated fine-tuning with an intermediate task before the target task ~\cite{phang2018sentence, wang2018glue, wang2018can, vu2020exploring, pruksachatkun2020intermediate}. The intermediate task is designed to accumulate additional knowledge to help the downstream target task achieve superior performance compared to a pre-trained model that is directly fine-tuned on the target task dataset. However, previous NLP studies~\cite{phang2018sentence} have demonstrated that selecting intermediate tasks arbitrarily can result in \textit{negative transfer}, whereby the target task's performance is adversely affected~\cite{wang2018glue}. Therefore, researchers have investigated various factors, including the learned linguistic skills~\cite{pruksachatkun2020intermediate}, dataset size, and data text domain~\cite{vu2020exploring}, to identify the conditions for \textit{positive transfer}, referring to a transfer learning between the intermediate task and the target task that leads to performance gains for the target task.



\begin{figure}[t]
    \centering
    \includegraphics[width=0.8\columnwidth]{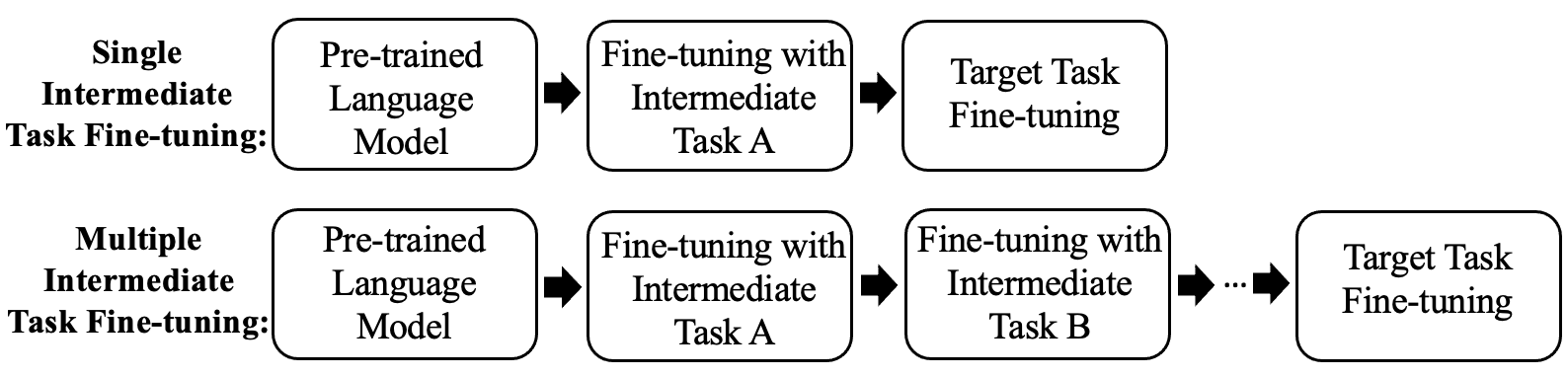}
    \caption{Single Intermediate Task Fine-tuning VS multiple Intermediate Task Fine-tuning}
    \label{fig:multiplefinetune}
\end{figure}

Despite these studies providing a foundational understanding of transferability between downstream tasks and the conditions for a positive transfer, they only considered a single intermediate task fine-tuned before the target task. In Figure \ref{fig:multiplefinetune}, we present the difference between single intermediate task fine-tuning and multiple intermediate task fine-tuning. We posit that fine-tuning more than one intermediate task in some specific orders may further improve the model's performance as the model may accumulate more diverse knowledge from multiple task-specific datasets. However, the influence of fine-tuning multiple intermediate tasks on the target task's performance has not been explored. Identifying the conditions for positive transfer in multiple intermediate task fine-tuning could be a way to overcome data scarcity for tasks with limited data availability. Additionally, 
prior research shows that training in multiple programming languages makes a model more generalizable~\cite{ahmed2022multilingual}, which lets us posit that
by fine-tuning through multiple intermediate tasks, a model can learn diverse features and patterns from different tasks and datasets. This exposure to diverse tasks can make the model more generalizable. Multiple intermediate fine-tuning can also mitigate the forgetting problem in deep learning models, which refers to the tendency of models to lose previously acquired knowledge due to newly fine-tuned tasks~\cite{kirkpatrick2017overcoming}. 
Therefore, in contrast to prior studies where only a single intermediate task was considered, we investigate whether refinement through multiple intermediate tasks helps.


Following the success of pre-trained large language models in NLP, language models pre-trained on a large corpus of source code such as CodeBERT~\cite{feng2020codebert}, PLBART~\cite{ahmad2021unified}, and CodeT5~\cite{wang2021codet5} have demonstrated remarkable performances in a variety of code intelligence tasks, spanning from code understanding (e.g., defect detection) to code generation (e.g., code translation). As source code possesses distinct features different from natural languages, such as its inherent syntactic constraints and semantic structures~\cite{casalnuovo2019studying, niu2023empirical}, the findings from NLP are not directly applicable to Software Engineering (SE) tasks that primarily involve source code. The transferability between SE tasks remains an area requiring further investigation. 


To shed light on SE task transferability and how SE tasks affect each other, we ask our first research question: 

\textbf{RQ1: Does the fine-tuning task ordering with multiple intermediate SE tasks matter for target task performance?}

To answer our RQ1, we select four SE tasks from CodeXGLUE benchmark suite~\cite{lu2021codexglue}, namely, Clone Detection (CD)~\cite{svajlenko2014towards}, Defect Detection (DD)~\cite{zhou2019devign}, Code Repair (CR)~\cite{tufano2019empirical}, and Code Translation (CT)~\cite{nguyen2015divide}. It should be noted that the selection of the four tasks is driven by the aim of exploring tasks that involve various methods of utilizing and manipulating code. Specifically, our investigation includes tasks that focus on (i) transforming code with the goal of changing its behavior (CR) and preserving its behavior (CT); (ii) understanding the code as to whether it contains a bug (DD); and (iii) understanding the code as to whether it is semantically similar to another piece of code (CD). Such a group of tasks contributes to increasing the generalizability of our findings. 

We picked CodeBERT~\cite{feng2020codebert} as the pre-trained model for analyzing since it has been widely evaluated and compared in a large number of studies~\cite{kanade2020learning,lu2021codexglue,wang2021codet5}. 
In our experiments, we first fine-tune each of the four tasks directly without any intermediate tasks in between to obtain the \textit{baseline models}. Then, we conduct the fine-tuning on SE tasks sequentially to collect \textit{fine-tuning chain models}. We fine-tune all permutations of any two, three, or all four tasks in sequential chains. That is, the chains' lengths range from two to lengths of four (e.g., two, three, or four tasks trained sequentially in a fine-tuning chain) that involve any possible permutations of two, three, or four tasks. We use \textit{task ordering} to refer to a sequential fine-tuning chain of intermediate and target tasks with the pre-trained model. All the permutations of tasks are fine-tuned sequentially with the goal of discovering task ordering patterns that either always bring performance gain or loss for a target task (i.e., positive/negative transfer), which may indicate the fine-tuning task ordering does matter for the target task. 


Next, we are interested in explaining why task ordering matters. This prompts us to ask our second research question:

\textbf{RQ2: Why does the fine-tuning task ordering impact the performance of the target task?}

To answer RQ2, we provide explanations from a variety of dimensions, including characteristics of training dataset, SE task, and model. In each dimension, we conduct experiments to investigate their associations with the impact of task orderings.



Finally, we analyze the practical implications of our findings and ask our third research question:

\textbf{RQ3: What are the time-performance trade-offs when choosing different task orderings to fine-tune?}
To answer RQ3, we analyze the fine-tuning chain models of positive transfer and their training time to find out whether certain trade-offs between training time and performance gain exist. Then, we provide insights into selecting the most cost-effective task ordering for a given target task.

To summarize, the significance of our contributions are following:

(1) We show that different task orderings could impact the target task performance differently.

(2) We analyze the possible factors in data, task, and model dimension that could help to explain why task ordering matters.

(3) We analyze the training trade-offs of task orderings to provide implications for practitioners and researchers.

\section{Related Works}
\label{sec:relatedworks}

\subsection{Fine-tuning of Pre-trained Language Models for SE tasks}
In the Pre-trained Language Model paradigm, rather than training the models from scratch with randomly initialized weights, the models are \textit{pre-trained} on large text corpora so that they learn the general knowledge about the target natural language~\cite{dai2015semi, radford2018improving}. Then the learned knowledge can be transferred to the downstream tasks by fine-tuning the pre-trained models with a relatively small amount of task-specific data~\cite{lu2021codexglue}. This pre-training approach has proven effective in improving the models' performance compared to the traditional approach~\cite{lu2021codexglue, chen2021evaluating}.

Pre-trained language models can be classified into autoregressive language models~\cite{brown2020language}, masked language models~\cite{devlin2018bert}, and encoder-decoder~\cite{lewis2019bart} language models based on their pre-training objectives. Autoregressive language models employ "generative pre-training", predicting the next token in a sequence (e.g., GPT-3)~\cite{brown2020language, recchia2021teaching}. Masked language models predict masked tokens using the remaining token sequence as context (e.g., BERT)~\cite{devlin2018bert}, while encoder-decoder language models are trained on sequence-to-sequence tasks, excelling in translation and summarization tasks (e.g., T5, BART)~\cite{lewis2019bart, raffel2020exploring}. Recently, pre-trained natural language models have been adapted for source code, such as CodeGPT/Codex~\cite{chen2021evaluating} (based on GPT-3~\cite{brown2020language}), CodeBERT/GraphCodeBERT~\cite{feng2020codebert, guo2020graphcodebert} (based on BERT~\cite{devlin2018bert, liu2019roberta}), and CodeT5~\cite{wang2021codet5} (based on T5~\cite{raffel2020exploring}). These language models have been evaluated and fine-tuned for various SE tasks demonstrating promising results. 

\subsection{Transfer Learning through Intermediate Task fine-tuning}
To further enhance state-of-the-art model performance on downstream NLP tasks, researchers have introduced an additional fine-tuning strategy: first fine-tuning the pre-trained model on an intermediate task before the target task of interest.~\cite{phang2018sentence, wang2018can, wang2018glue}. However, this approach does not consistently improve target task performance and may even be detrimental, depending on the chosen intermediate task~\cite{wang2018can}. 

To investigate the condition of positive transfer and to avoid negative transfer~\cite{torrey2010transfer}, Pruksachatkun et al.~\cite{pruksachatkun2020intermediate} examined the linguistic skills acquired from intermediate tasks that could potentially contribute to positive transfer on a target task. Their findings suggested that although a model might learn certain skills from an intermediate task, these skills may not necessarily benefit the target task. Change et al. ~\cite{chang2021rethinking} investigated various factors that could influence the effectiveness of intermediate task fine-tuning, including the training dataset sizes of both intermediate and target tasks. Furthermore, Vu et al.~\cite{vu2020exploring} carried out extensive experiments on 33 NLP tasks to illuminate the transferability between these tasks, discovering that the similarity between the intermediate and target tasks, as well as their training data domains, are important factors for positive transfer. 

Despite these previous studies providing valuable insights into transferability among different downstream tasks, they only focused on NLP tasks, not SE tasks, which primarily involve program source code instead of natural language, making it a non-trivial task to adapt learning from NLP tasks to SE tasks. Consequently, as a community, we lack an understanding of transferability between SE tasks. Moreover, these NLP studies only considered a single intermediate task, neglecting the potential effects of multiple intermediate tasks and their fine-tuning orderings on the target task. It is worth noting that Mastropaolo et al.~\cite{mastropaolo2022using} explored the advantages of transfer learning for SE tasks, but their investigation was limited to multi-task learning with only four SE tasks (muti-task learning aims to train tasks in parallel using a shared representation, while our approach is training tasks sequentially in chains), leaving the impacts of sequential intermediate-target task fine-tuning unexplored. In this paper, we take the first steps towards exploring that.

\section{Experimental Setup}
\label{sec:method}

\begin{figure}[htp]
    \centering
    \includegraphics[width=0.3\columnwidth]{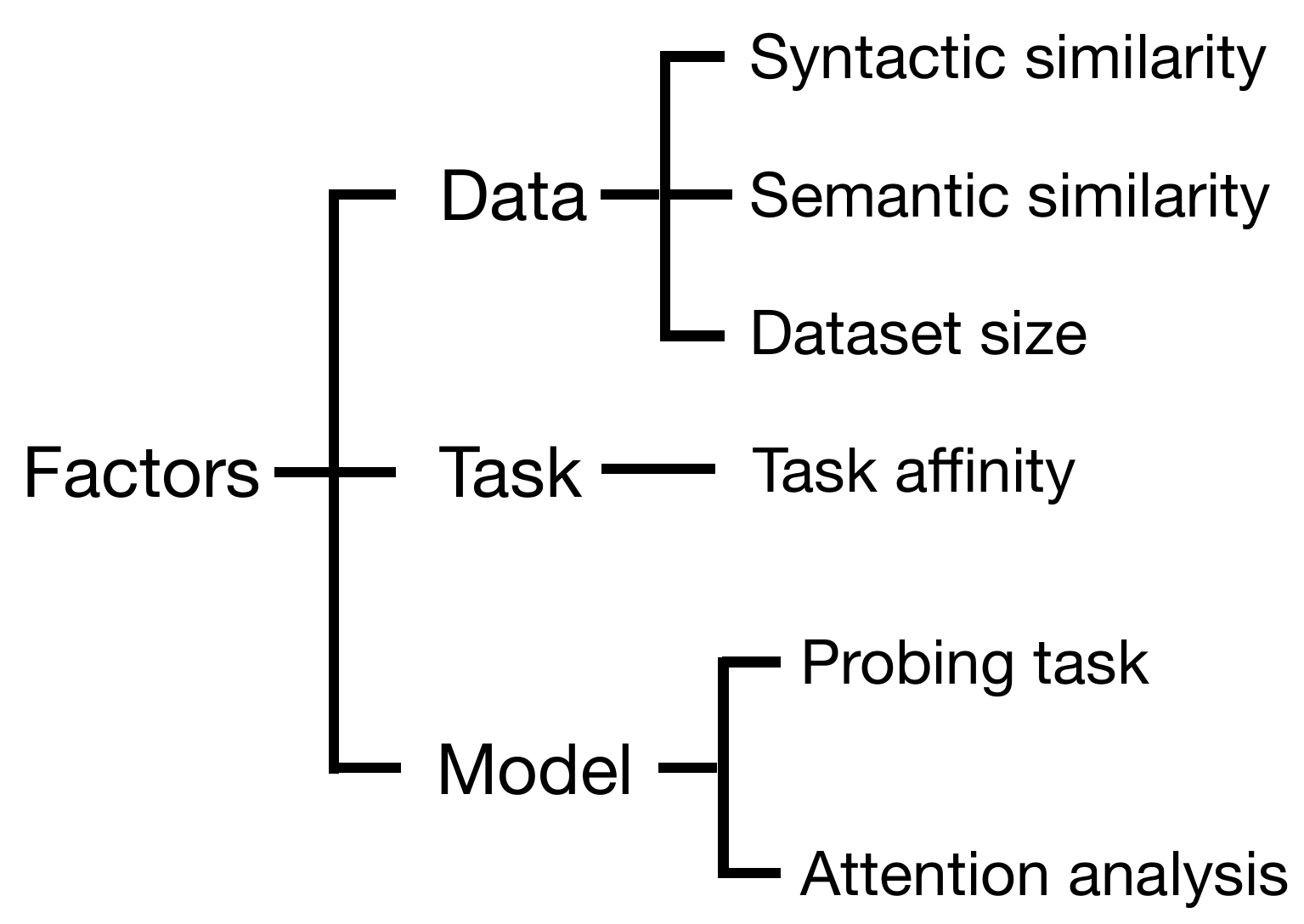}
    \caption{Overview of Factors Explored in this Study}
    \label{fig:overview}
\end{figure}

Our goal was to investigate whether the fine-tuning task orderings of SE tasks have an impact on the performance of the target task (RQ1). Then, we aimed to explore the possible reasons from various dimensions for such an impact (RQ2). Figure \ref{fig:overview} shows a structural view of all the factors we analyzed. Lastly, we aimed to give suggestions for selecting the most cost-effective task orderings for a given target task (RQ3). In the following subsections, we detail the applied experimental setup to answer our research questions. 

\subsection{Experimental Pipeline}

In this study, we select four downstream SE tasks (CD, DD, CR, and CT) from CodeXGLUE~\cite{lu2021codexglue}, a benchmark dataset for code intelligence. Table \ref{tab:tasks} enumerates the four SE tasks we used in our experiments. We list brief definitions of these selected tasks and datasets below:

\noindent{\textbf{Clone Detection (CD)}}
is a task of retrieving similar code given a code snippet as a query. For CD, we use POJ-104 dataset consisting of 52k examples.

\noindent\textbf{Defect Detection (DD)}
is to identify whether a given function is vulnerable or not, which is a binary classification task. We use Devign dataset consisting of 26.4k examples.

\noindent\textbf{Code Repair (CR)}
aims to automatically repair bugs in the code. We use Bugs2Fix dataset with 122k examples for CR.

\noindent\textbf{Code Translation (CT)}
aims to translate source code from one programming language to another programming language. In this study, the translation is between Java and C\# as it is the only pair provided in the CodeXGLUE dataset. For CT, we use CodeTrans dataset consisting of 11.5k examples.

All four tasks are among the widely-used SE tasks for evaluating pre-trained models of source code~\cite{niu2023empirical}. Table \ref{tab:tasks} shows the abbreviation (Ab.), the dataset, and the main evaluation metric. The evaluation metrics and datasets are consistent with those used by CodeXGLUE for each corresponding task. 
The objective of our study is to investigate the impact of intermediate task orderings on the performance of different target tasks. We utilize varying numbers of SE tasks within fine-tuning chains, where one to three tasks function as intermediate tasks, sequentially fine-tuned before the target task. This results in a total of 60 fine-tuning chain models, with lengths (i.e., the total number of fine-tuning tasks in a chain) varying from 2-4. To ensure comprehensive evaluation, we employ a 10-fold cross-validation technique on a combined dataset of train, dev, and test data from the corresponding task in CodeXGLUE. We establish baseline models for the four tasks by training directly with pre-trained model on the target task datasets. This enables us to determine the statistical significance of any performance disparities between a fine-tuning chain model and a baseline model, facilitating the identification of interesting task ordering patterns. Other settings, including hyper-parameters and the optimizer, are the same as CodeXGLUE.



Our experiments utilize CodeBERT as the pre-trained encoder for all four tasks. We select CodeBERT for two primary reasons: firstly, it was the state-of-the-art model when our research embarked, and secondly, it has been evaluated and compared in many studies~\cite{kanade2020learning,lu2021codexglue,wang2021codet5}. We did not try pre-trained models of source code other than CodeBERT as we believe that all models have similar underlying foundational structure based on transformer architecture. Even though observations learnt through analyzing one model may not be exactly identical to other models, it is safe to assume that the patterns discovered are going to be homogeneous due to similar structure. We follow a sequential fine-tuning approach, whereby the encoder is fine-tuned on each intermediate task before proceeding to the target task. For instance, in a task order consisting of four tasks, the encoder is fine-tuned on three intermediate tasks in sequence before being fine-tuned with the target task's dataset. Each task comprises its own unique component or layers. For code generation tasks (CR and CT), we follow CodeXGLUE's setup by employing a randomly initialized Transformer with six layers, 768-dimensional hidden states, and 12 attention heads as the decoder. In the case of CR and CT being part of the same fine-tuning chain, the decoder fine-tuned on one task serves as the starting point for the subsequent task, enabling it to leverage the knowledge accumulated from the previous task. We use softmax layers as task-specific components for classification (DD) and retrieval (CD).



To keep the number of experiments manageable, we did not try SE tasks other than the four we picked. We did not find it reasonable to try more tasks since we can simply expect that the number of experiments would increase to an unaffordable level, which would be computationally expensive and time-consuming. We believe this selection is sufficient for us to analyze and explain the general impact of different task orderings on target task performance. In addition, our goal is not to identify the optimal task ordering for a target task fine-tuned with a specific pre-trained model but to investigate task orderings' impacts and the potential factors that may help explain.


\begin{table}[t]
\centering
\caption{Details of Evaluation Tasks, Datasets, and Metrics}
\label{tab:tasks}
\resizebox{0.7\columnwidth}{!}{%
\begin{tabular}{|c|c|c|c|c|c|}
\hline
\textbf{Task}    & \textbf{Ab,} & \textbf{Dataset} & \textbf{Dataset Size} & \textbf{Language} & \textbf{Metric} \\ \hline
Defect Detection & DD           & Devign           & 26.4k                 & C                 & Accuracy        \\ \hline
Clone Detection  & CD           & POJ-104          & 52k                   & C/C++             & MAP@R           \\ \hline
Code Repair      & CR           & Bugs2Fix         & 122k                  & Java              & BLEU            \\ \hline
Code Translation & CT           & CodeTrans        & 11.5k                 & Java-C\#          & BLEU            \\ \hline
\end{tabular}%
}
\end{table}

\subsection{Analyses of Dataset Characteristics}

Next, we analyze various characteristics of the training dataset in order to explain the impact on performance. The characteristics we selected have been used in NLP research for analyzing the positive transfer learning between a variety of downstream tasks~\cite{vu2020exploring,chang2021rethinking}.

\noindent\textit{\textbf{Syntactic Similarity Analysis: }} Ahmed et al.~\cite{ahmed2022multilingual} have shown that the language model's performance on downstream SE tasks can be improved by utilizing shared syntactic information between datasets consisting of code snippets with similar purposes but written in different programming languages. Motivated by this, we explore the syntactic similarities of datasets by focusing on three code-related characteristics: \textit{keywords}, \textit{operators}, and \textit{identifiers}. However, analyzing all possible pairs of programs from each dataset pair would be unfeasible due to the large size of each dataset. Thus, we choose to utilize a sampling approach to reduce the size. We determine the appropriate sample size for each dataset with a confidence level of 99\% and a margin of error of 2\% and then proceed to analyze the resulting sampled dataset.

We calculate the similarity following the procedure below:

(1) We randomly sample datasets using the calculated sample size. 

(2) Every dataset consists of code snippets. We go through the sampled dataset and gather information on the keywords, operators, and identifiers of each code snippet.

(3) We loop through the combination that contains pairs of code snippets selected from each pair of datasets. For every pair of code snippet, we apply the Jaccard similarity formula (J = $ \frac{|A \cap B|}  {|A \cup B|} $ where A and B are two sets)~\cite{niwattanakul2013using} to calculate their similarities in terms of the keywords, operators, and identifiers. 

 (4) After computing the similarity scores for each pair of code snippets, we calculate the overall similarity of the two datasets by taking the average similarity across all code snippets.

\noindent\textit{\textbf{Semantic Similarity Analysis: }}
After examining the syntactic similarities,  we shift our attention to analyzing semantic similarity among datasets. To extract the semantic information of a code snippet, we conduct feature extraction using CodeBERT since the pre-trained CodeBERT model can provide a reliable vector representation containing the code snippets' semantic information~\cite{feng2020codebert}.

Our approach involves employing the CodeBERT model to obtain vector representations for code snippets. We then evaluate the semantic similarity between the pair of code snippets from two datasets by computing the cosine similarity~\cite{rahutomo2012semantic}. We apply cosine similarity in semantic similarity analysis since prior research showed that the vectors used for measuring cosine similarity capture a substantial amount of semantic information~\cite{mikolov2013exploiting}. Ultimately, we utilize the similarity scores obtained from all pairs of code snippets from both datasets to determine the semantic similarity between the two datasets. Specifically, we average the cosine similarities for all pairs of code snippets from the two datasets to calculate the semantic similarity of the two datasets.

Due to the enormous size of each dataset, analyzing all possible pairs of code snippets from each dataset would be a daunting task. Thus, we resort to sampling. To determine the appropriate sample size for each dataset, we utilize the same sample size calculator~\cite{Sample-size-calculator} with a 99\% confidence level and a 2\% margin of error. Subsequently, we sample each dataset using the population size calculated by the sample size calculator. Our analysis steps are the following:

(1) We select two datasets and then proceed to sample them using the previously calculated sample size.

(2) For each pair of datasets, we loop through their combinations which is a set of pairs of code snippets. 

(3) To generate vector representations for both code snippets, we use CodeBERT. However, CodeBERT has a constraint on the number of input tokens which is 512. For code snippets with more than 512 tokens, we split them into multiple parts, each containing a maximum of 512 tokens. After that, we construct vector representations for each part and concatenate them all. This way, we get the vector representation of the code snippet. 

(4) We determine the semantic similarity between the two code snippets using the cosine similarity between their vector representations. We repeat this step for every pair of code snippets in the combination. We then determine the semantic similarity of the two datasets by averaging the cosine similarity scores across all pairs of code snippets.

(5) We repeat the previously described steps for all pairs of datasets.

\noindent\textit{\textbf{Dataset Size Analysis:}} Many research works in NLP have explored the influence of both intermediate and target dataset sizes on the performance of target tasks. These investigations found that target tasks with smaller datasets gained the most from intermediate task training in general. However, they discovered no significant, consistent effect of intermediate task dataset size on the performance of target tasks. Instead, they highlighted the importance of other factors for positive transfer between certain NLP tasks~\cite{phang2018sentence, vu2020exploring, pruksachatkun2020intermediate, chang2021rethinking}. In our study, we examined the potential role of training dataset size in explaining the varying impacts of specific fine-tuning task orderings on target task performance. 
\subsection{Analyses of SE Task Characteristics}
\label{sec:SE task sim}
\noindent\textit{\textbf{Task Affinity Analysis:}}
Based on the study by Fifty et al.~\cite{fifty2021efficiently}, the level of task affinity between two tasks can reveal the extent to which one task can benefit another task. In other words, it can indicate how much knowledge can be transferred between the two tasks. Given that our training approach updates weights based on the SE tasks in the task orderings, we posit that task affinity is a crucial factor in determining the models' performance.


The aim of this analysis is to investigate the task affinity between tasks. We define \textit{intermediate fine-tuned model}: The fine-tuning chain model where the pre-trained model is fine-tuned with the intermediate tasks in the task orderings without the target task, which is the starting point for the target task to train on. For example, for task ordering CR$\rightarrow$CT, intermediate fine-tuned model would be the CR baseline model.


We start by training the pre-trained, baseline, and intermediate fine-tuned models on the target task, and record their losses. Then, we calculate the sum of the differences in losses between the intermediate fine-tuned and baseline models, which we label as diff\_intermediate. Similarly, we calculate the sum of the loss differences between the pre-trained and baseline models, known as diff\_pre-trained. As an example, suppose the task ordering is CR$\rightarrow$CT, the intermediate fine-tuned model would be the baseline CR model and the target task would be CT. We train the pre-trained model with CT dataset, record the losses and label it as pre-trained\_loss. We then train the intermediate fine-tuned model, which has been trained on any of the tasks besides CT with CT dataset, record the losses and label it as intermediate\_loss. We lastly train the CT baseline model, which has been first trained on CT again with CT dataset, record the losses and label it as baseline\_loss. Thus, the diff\_intermediate equals sum of differences between intermediate\_loss and baseline\_loss. The diff\_pre-trained equals sum of differences between pre-trained\_loss and baseline\_loss.

We then compare diff\_intermediate with diff\_pre-trained to determine the effectiveness of knowledge transfer from intermediate tasks to the target task, which is the task affinity. We should note that this analysis is asymmetric because, for any task ordering, we compare the intermediate fine-tuned model with its baseline model. We consider this method valid because we use the baseline model's losses on the target task as a reference point for comparison. Compared to pre-trained model and intermediate fine-tuned model, baseline model is the only model who has been trained on the target task. Therefore, when retraining it on the target task, its losses on the target task would be the ideal case, and by comparing diff\_pre-trained and diff\_intermediate, we would tell the model that is closer to the baseline model, which explains whether the intermediate tasks help transfer knowledge.

\subsection{Analyses of Model Characteristics}
\label{sec:modelanalysis}

\noindent\textit{\textbf{Probing Tasks Analysis: }}Having a good performance on an SE task usually requires the model have a good understanding of some code properties. In this study, we posit that models fine-tuned with different task-specific datasets may exhibit different extents of understandings of a certain code property, and the fine-tuning of some intermediate tasks can affect the target task's learning of a code property critical for the model to perform well on  target task.

To evaluate the acquisition of language skills learned by a model, researchers have come up with \textit{probing tasks}. Probing tasks are tasks that are designed to understand what linguistic property is encoded in pre-trained language models. These tasks are diagnostic in nature, where a simple linear classifier is trained on the vector embeddings of a language model to predict specific properties of its input (i.e. natural language texts). The performance of the classifier on probing tasks indicates whether probed information exists in the vector embedding of a model and the extent to which such information is encoded or learned. Probing has been extensively designed and studied for models trained with natural language with the aim of analyzing a large amount of natural language properties and skills~\cite{adi2016fine, alain2016understanding, belinkov2017neural, conneau2018you, tenney2019you}.

Recently, probing tasks from NLP have been adapted to pre-trained language models of source code. Various probing tasks have been constructed to evaluate the hidden vectors of a model and their ability to grasp code-related characteristics after model pre-training stage~\cite{karmakar2021pre, troshin2022probing, wan2022they}. These code-related characteristics may be useful in downstream SE tasks. In our study, we use two of the probing tasks from \cite{karmakar2021pre} to assess the models' learning degrees of source code semantic and syntactic information. The probing task that is designed for testing the model's syntactic understanding of source code is Abstract Syntax Tree Node Tagging (AST), which is a task of identifying AST node tags for a sequence of code tokens. Another one that is for assessing the model's awareness of code semantics is Invalid Type Detection (TYP). The model is asked to distinguish between code snippets that contain misspelled primitive data types from semantically valid ones. 

In this study, we train a logistic regression classifier~\cite{lavalley2008logistic} (the hyper-parameters are set the same as those from \cite{karmakar2021pre}) that takes the input feature vectors from the frozen hidden layers of fine-tuning chain models. The accuracies are based on the vector embeddings from each of the hidden layers in CodeBERT ranging from 1-12. Pre-trained CodeBERT has been probed to have heterogeneous probing task performance across layers~\cite{karmakar2021pre}. To evaluate the general understanding of a particular probed skill from a fine-tuned model, we take the average of all 12 accuracies (from all 12 layers) as an estimate of the model's overall performance as we believe that learned knowledge of a code property is spread across all layers. We call this average the \textit{probing task performance} for a fine-tuned model in the following sections. Next, we investigated the relationship between probing and target tasks in order to see whether a probed code-related skill is important for a fine-tuned model to perform well on this target task. The \textit{target task performance} is the average of the performances of all 10-fold cross validation models. Similar to \cite{pruksachatkun2020intermediate}, we calculate the Spearman correlation~\cite{myers2004spearman} between probing task and target task performances for all fine-tuning chain models whose task orderings end with the target task. With the knowledge of different SE tasks' requirements for the probed skills, we finally provide insights into why certain task orderings boost or hurt the target task performance (i.e. positive transfer or negative transfer) by comparing the performances of probing tasks.

\noindent\textit{\textbf{Attention Analysis: }}In this study, we investigate the attention weight assignments of CodeBERT's heads on different syntax tokens in its input sequence. CodeBERT is a transformer-based model that generates self-attention weights for each token in the input sequence. These weights are an indicator of the level of attention that other tokens in the input sequence give to each token. CodeBERT consists of 12 self-attention layers, with each layer having 12 heads that produce attention weights for the same token. To obtain a comprehensive measure of the attention weight for each token, we average the attention weights across all layers and heads.

To analyze the attention weights assigned by CodeBERT's heads, we consider different versions of the fine-tuned model and assess the salience of various syntax tokens, including identifiers, modifiers, operators, data types, separators, keywords, strings, and Booleans. These syntax tokens were selected based on prior work by Aljehane et al.~\cite{aljehane2021determining}. We extract these syntax tokens from the input source code using Javalang~\cite{javalang}, a widely used Java syntax collection library. Additionally, we investigate the role of abstract syntax structures, such as method signature, if-else statements, while statements, and return statements, in determining the attention behavior of fine-tuned CodeBERT. To identify these abstract syntax tree structures, we utilize tree-sitter-java~\cite{tree-sitter}, a commonly used Java abstract syntax tree structure identification library.

\subsection{Analysis of Time and Performance}
\label{sec:timeanalysis}
Training deep learning models on SE tasks consumes relatively large amount of hardware and time resources.~\cite{strubell2020energy, strubell2019energy}. This training time can vary from a few hours to hundreds of hours. When selecting the task ordering for a given target task, the performance should not be the only factor to consider, but also the resources it consumes. 

In this analysis, we aim to answer RQ3, which focuses on considering the time-performance trade-offs to pick the optimal task ordering for performance improvement. We examine all the task orderings of positive transfer and compare the model performance gain with the additional training time required. Specifically, we estimate the model performance gain by using \textit{relative performance gain} from~\cite{vu2020exploring}, which is defined as $$g_{fine-tuning\_chain\_model} = \frac{P_{fine-tuning\_chain\_model} - P_{baseline}}{P_{baseline}}$$  where $P_{baseline}$ represents baseline model performance and\\ 
$P_{fine-tuning\_chain\_model}$ represents the performance of the fine-tuning chain model with the same target task. Furthermore, in order to  analyze the cost-effectiveness of each task ordering, we define \textit{effective ratio} as


$$effective\_ratio = \frac{g_{fine-tuning\_chain\_model}}{\mathit{\Delta}training\_time} \; $$

Where ${\mathit{\Delta}training\_time}$ represents the training time difference between baseline model and the fine-tuning chain model with the same target task. The effective ratio provides an intuitive way of understanding the cost of performance improvement, namely performance improvement per unit of time.

\section{Evaluation Results of Model Performances}
\label{sec:results}

In order to answer our RQ1 about whether task ordering matters, we fine-tune all permutations of four tasks sequentially on CodeBERT. For each fine-tuning chain model whose task ordering ends with a target task, we report its metric score and compare the score with the performance of its baseline model. It is worth noting that we only compare the performance of fine-tuning chain model with the same target task for fair comparison. 
Given that there are four tasks, the total number of chain models with varying lengths of task orderings is 60. Since we trained the models with 10-fold cross validation, we conduct Welch’s t-test~\cite{ruxton2006unequal} to measure the statistical significance of the performance difference between fine-tuning chain models and baselines to identify whether the chain model is a positive or negative transfer. We show the results in Table \ref{tab:chain-performances}. We also highlight the positive transfer and negative transfer for each model. Based on our results, several interesting patterns emerge. For example, the performances of CD and DD almost always dropped by intermediate task fine-tuning no matter what the intermediate tasks are, while the different task orderings of CR and CT could bring either performance gain or loss. Due to space constraints, in this paper we select two patterns which can be distilled from highlighted entries in Code translation and Code refinement columns of Table \ref{tab:chain-performances}. These patterns can trigger both positive and negative transfer depending on the task orderings.


\textbf{Pattern 1 (positive transfer ``+'')} When CR is fine-tuned before CT (CR$\rightarrow$CT), CT’s performance becomes statistically significantly \textit{better} than CT baseline. Other intermediate tasks that are fine-tuned between CR and CT in the task orderings of fine-tuning chain models \textit{do not} affect this positive transfer and statistical significance.

\textbf{Pattern 2 (negative transfer``-'')} When CT is fine-tuned before CR (CT$\rightarrow$CR ), CR’s performance becomes statistically significantly \textit{worse} than CR baseline. Other intermediate tasks that are fine-tuned between CT and CR \textit{do not} affect this negative transfer and statistical significance.

\begin{table*}[]
\centering
\caption{The Performances of Fine-tuning Chain Models and Baselines (``+'' denotes positive transfer while ``-'' denotes negative transfer to the target tasks)}
\label{tab:chain-performances}
\resizebox{0.95\linewidth}{!}{%
\begin{tabular}{|lc|lc|lc|lc|}
\hline
\multicolumn{2}{|c|}{\textbf{Code translation (CT)}} & \multicolumn{2}{c|}{\textbf{Code refinement (CR)}} & \multicolumn{2}{c|}{\textbf{Clone detection (CD)}} & \multicolumn{2}{c|}{\textbf{Defect defection (DD)}} \\ \hline
\multicolumn{1}{|c|}{Models} & \begin{tabular}[c]{@{}c@{}}Average \\ BLEU Scores\end{tabular} & \multicolumn{1}{c|}{Models} & \begin{tabular}[c]{@{}c@{}}Average \\ BLEU Scores\end{tabular} & \multicolumn{1}{c|}{Models} & \begin{tabular}[c]{@{}c@{}}Average \\ MAP@R Score\end{tabular} & \multicolumn{1}{c|}{Models} & \begin{tabular}[c]{@{}c@{}}Average\\ Accuracy\end{tabular} \\ \hline
\multicolumn{1}{|l|}{CT}              & 70.881                       & \multicolumn{1}{l|}{CR}              & 79.566                       & \multicolumn{1}{l|}{CD}              & 0.869                         & \multicolumn{1}{l|}{DD}              & 0.630                       \\ \hline
\multicolumn{1}{|l|}{CR$\rightarrow$CT}           & \textbf{74.242 (+)}          & \multicolumn{1}{l|}{CT$\rightarrow$CR}           & \textbf{76.824 (-)}          & \multicolumn{1}{l|}{CT$\rightarrow$CD}           & 0.871                         & \multicolumn{1}{l|}{CR$\rightarrow$DD}           & 0.626                       \\ \hline
\multicolumn{1}{|l|}{CD$\rightarrow$CT}           & 70.879                       & \multicolumn{1}{l|}{CD$\rightarrow$CR}           & 79.643                       & \multicolumn{1}{l|}{CR$\rightarrow$CD}           & \textbf{0.853 (-)}            & \multicolumn{1}{l|}{CD$\rightarrow$DD}           & \textbf{0.619 (-)}          \\ \hline
\multicolumn{1}{|l|}{DD$\rightarrow$CT}           & 70.839                       & \multicolumn{1}{l|}{DD$\rightarrow$CR}           & 79.712                       & \multicolumn{1}{l|}{DD$\rightarrow$CD}           & \textbf{0.853 (-)}            & \multicolumn{1}{l|}{CT$\rightarrow$DD}           & 0.628                       \\ \hline
\multicolumn{1}{|l|}{CR$\rightarrow$CD$\rightarrow$CT}        & \textbf{73.957 (+)}          & \multicolumn{1}{l|}{CT$\rightarrow$CD$\rightarrow$CR}        & \textbf{77.870 (-)}          & \multicolumn{1}{l|}{CT$\rightarrow$CR$\rightarrow$CD}        & \textbf{0.847 (-)}            & \multicolumn{1}{l|}{CR$\rightarrow$CD$\rightarrow$DD}        & \textbf{0.610 (-)}          \\ \hline
\multicolumn{1}{|l|}{CD$\rightarrow$CR$\rightarrow$CT}        & \textbf{75.058 (+)}          & \multicolumn{1}{l|}{CD$\rightarrow$CT$\rightarrow$CR}        & \textbf{77.455 (-)}          & \multicolumn{1}{l|}{CR$\rightarrow$CT$\rightarrow$CD}        & 0.862                         & \multicolumn{1}{l|}{CD$\rightarrow$CR$\rightarrow$DD}        & 0.626                       \\ \hline
\multicolumn{1}{|l|}{DD$\rightarrow$CD$\rightarrow$CT}        & 70.739                       & \multicolumn{1}{l|}{DD$\rightarrow$CD$\rightarrow$CR}        & 79.825                       & \multicolumn{1}{l|}{DD$\rightarrow$CR$\rightarrow$CD}        & \textbf{0.855 (-)}            & \multicolumn{1}{l|}{CT$\rightarrow$CD$\rightarrow$DD}        & \textbf{0.610 (-)}          \\ \hline
\multicolumn{1}{|l|}{CD$\rightarrow$DD$\rightarrow$CT}        & 71.000                       & \multicolumn{1}{l|}{CD$\rightarrow$DD$\rightarrow$CR}        & 79.616                       & \multicolumn{1}{l|}{CR$\rightarrow$DD$\rightarrow$CD}        & \textbf{0.854 (-)}            & \multicolumn{1}{l|}{CD$\rightarrow$CT$\rightarrow$DD}        & 0.628                       \\ \hline
\multicolumn{1}{|l|}{DD$\rightarrow$CR$\rightarrow$CT}        & \textbf{74.113 (+)}          & \multicolumn{1}{l|}{DD$\rightarrow$CT$\rightarrow$CR}        & \textbf{77.347 (-)}          & \multicolumn{1}{l|}{DD$\rightarrow$CT$\rightarrow$CD}        & \textbf{0.855 (-)}            & \multicolumn{1}{l|}{CT$\rightarrow$CR$\rightarrow$DD}        & 0.621                       \\ \hline
\multicolumn{1}{|l|}{CR$\rightarrow$DD$\rightarrow$CT}        & \textbf{74.118 (+)}          & \multicolumn{1}{l|}{CT$\rightarrow$DD$\rightarrow$CR}        & \textbf{78.029 (-)}          & \multicolumn{1}{l|}{CT$\rightarrow$DD$\rightarrow$CD}        & \textbf{0.866 (-)}            & \multicolumn{1}{l|}{CR$\rightarrow$CT$\rightarrow$DD}        & 0.627                       \\ \hline
\multicolumn{1}{|l|}{DD$\rightarrow$CR$\rightarrow$CD$\rightarrow$CT}     & \textbf{73.942 (+)}          & \multicolumn{1}{l|}{DD$\rightarrow$CT$\rightarrow$CD$\rightarrow$CR}     & \textbf{77.936 (-)}          & \multicolumn{1}{l|}{DD$\rightarrow$CT$\rightarrow$CR$\rightarrow$CD}     & \textbf{0.844 (-)}            & \multicolumn{1}{l|}{CT$\rightarrow$CR$\rightarrow$CD$\rightarrow$DD}     & \textbf{0.605 (-)}          \\ \hline
\multicolumn{1}{|l|}{DD$\rightarrow$CD$\rightarrow$CR$\rightarrow$CT}     & \textbf{73.950 (+)}          & \multicolumn{1}{l|}{DD$\rightarrow$CD$\rightarrow$CT$\rightarrow$CR}     & \textbf{77.279 (-)}          & \multicolumn{1}{l|}{DD$\rightarrow$CR$\rightarrow$CT$\rightarrow$CD}     & \textbf{0.853 (-)}            & \multicolumn{1}{l|}{CT$\rightarrow$CD$\rightarrow$CR$\rightarrow$DD}     & \textbf{0.614 (-)}          \\ \hline
\multicolumn{1}{|l|}{CR$\rightarrow$DD$\rightarrow$CD$\rightarrow$CT}     & \textbf{74.145 (+)}          & \multicolumn{1}{l|}{CT$\rightarrow$DD$\rightarrow$CD$\rightarrow$CR}     & \textbf{78.120 (-)}          & \multicolumn{1}{l|}{CT$\rightarrow$DD$\rightarrow$CR$\rightarrow$CD}     & \textbf{0.841 (-)}            & \multicolumn{1}{l|}{CR$\rightarrow$CT$\rightarrow$CD$\rightarrow$DD}     & \textbf{0.607 (-)}          \\ \hline
\multicolumn{1}{|l|}{CR$\rightarrow$CD$\rightarrow$DD$\rightarrow$CT}     & \textbf{73.882 (+)}          & \multicolumn{1}{l|}{CT$\rightarrow$CD$\rightarrow$DD$\rightarrow$CR}     & \textbf{78.076 (-)}          & \multicolumn{1}{l|}{CT$\rightarrow$CR$\rightarrow$DD$\rightarrow$CD}     & \textbf{0.854 (-)}            & \multicolumn{1}{l|}{CR$\rightarrow$CD$\rightarrow$CT$\rightarrow$DD}     & \textbf{0.619 (-)}          \\ \hline
\multicolumn{1}{|l|}{CD$\rightarrow$DD$\rightarrow$CR$\rightarrow$CT}     & \textbf{74.606 (+)}          & \multicolumn{1}{l|}{CD$\rightarrow$DD$\rightarrow$CT$\rightarrow$CR}     & \textbf{77.745 (-)}          & \multicolumn{1}{l|}{CR$\rightarrow$DD$\rightarrow$CT$\rightarrow$CD}     & \textbf{0.857 (-)}            & \multicolumn{1}{l|}{CD$\rightarrow$CT$\rightarrow$CR$\rightarrow$DD}     & \textbf{0.606 (-)}          \\ \hline
\multicolumn{1}{|l|}{CD$\rightarrow$CR$\rightarrow$DD$\rightarrow$CT}     & \textbf{74.204 (+)}          & \multicolumn{1}{l|}{CD$\rightarrow$CT$\rightarrow$DD$\rightarrow$CR}     & \textbf{77.770 (-)}          & \multicolumn{1}{l|}{CR$\rightarrow$CT$\rightarrow$DD$\rightarrow$CD}     & \textbf{0.856 (-)}            & \multicolumn{1}{l|}{CD$\rightarrow$CR$\rightarrow$CT$\rightarrow$DD}     & \textbf{0.615 (-)}          \\ \hline
\end{tabular}%
}
\end{table*}

\section{Analyses of Explanatory Factors}
\label{sec:discussion}

To determine the factors that may help explain the positive and negative transfer patterns we found, we conducted multiple analyses from three perspectives: dataset, task, and model (as shown in Figure 2). For \textbf{Pattern 1} and \textbf{Pattern 2}, the intermediate tasks fine-tuned in between do not affect the statistical significance of negative transfer/positive transfer. 
Based on these findings, we simplify our analyses by only focusing on CR$\rightarrow$CT, CT$\rightarrow$CR 
fine-tuning chain models. In this section, we report the results of each analysis and provide possible explanations based on the results.

\subsection{Data: Syntactical Similarity Analysis}
In this section, we provide the average similarity scores for the dataset pairs (CR and CT) and (CD and DD) across keywords, operators, and identifiers. Although the two patterns we discovered are CR$\rightarrow$CT, and CT$\rightarrow$CR, we report the results of both (CR and CT) datasets and (CD and DD) datasets because we want to have a reference point for comparison. Our results demonstrate that the CR and CT datasets have a higher similarity in terms of operators (0.342 for 12,676,329 pairs of code snippets in CR and CT compared to 0.295 for 12,836,220 pairs of code snippets in CD and DD). However, they exhibit lower similarity in terms of keywords (0.247 for 13,789,146 pairs of code snippets in CR and CT compared to 0.305 for 12,850,932 pairs of code snippets in CD and DD) and identifiers (0.006 for 13,789,146 pairs of code snippets in CR and CT compared to 0.025 for 12,858,288 pairs of code snippets in CD and DD).

The lower similarity in keywords and identifiers between the CR and CT datasets compared to the CD and DD datasets may be explained by the difference in programming languages used. Specifically, the former uses Java and C\#, while the latter uses C/C++. Since Java and C\# have distinct keywords whereas C and C++ have similar keywords, this difference in programming languages could explain the lower similarity in terms of keywords. Furthermore, our findings regarding identifiers do not correspond with the findings in ~\cite{ahmed2022multilingual}. We speculate that the differences in our findings regarding identifiers compared to ~\cite{ahmed2022multilingual} is likely due to the disparities in the datasets used. The study by ~\cite{ahmed2022multilingual} utilized the ROSETTACODE dataset~\cite{RosettaCodeData} which consists of a large number of problems. For each problem, the answer is provided by code snippets written in different languages, which makes their meanings semantically equivalent. In contrast, our datasets do not have access to code snippets with similar semantic meanings, which could be the reason for the differences in our conclusions regarding identifiers. As a conclusion, we suggest that the higher similarity in terms of operators used in CR and CT could potentially account for the positive transfer (\textbf{Pattern 1}), as discussed in Section \ref{sec:results}.


\observation{The shared operators within the CR and CT datasets could be a possible explanation for the positive transfer when CR is fine-tuned before CT}.

\subsection{Data: Semantic Similarity Analysis}


We present our results of semantic similarity analysis for task pairs (CD, DD) and (CR, CT) in this section. Although the two patterns we discovered are CR$\rightarrow$CT, and CT$\rightarrow$CR, we report the results of both (CR and CT) datasets and (CD and DD) datasets because we want to have a reference point for comparison purpose. We report our semantic similarity for each pair of datasets by using the average cosine similarity scores of all pairs of code snippets in that pair of datasets. Our results show that the average similarity score for CD and DD's training datasets is 0.317 for a total of 421,838 pairs of code snippets, whereas the average similarity score for CR and CT's training datasets is 0.410 for a total of 1,645,128 pairs of code snippets. Based on these results, we could see that the code snippets in CR are more semantically similar to those in CT datasets than code snippets in CD dataset to DD datasets. This result indicates that the semantic similarity of code snippets between training datasets of different tasks could be a relevant factor to consider when selecting multiple SE tasks as intermediate tasks to enhance model performance. 



\observation{Choosing datasets that are more semantically similar could lead to better performance.}

\subsection{Data: Dataset Size Impact Analysis}

Among all task-specific datasets, CT's task-specific dataset has the smallest number of samples (Table \ref{tab:tasks}). When CT is designated as the target task, its performance improves by intermediate task fine-tuning in most of the fine-tuning chain models. And it is the only task that shows positive transfer by intermediate task fine-tuning in our experiments. This observation aligns with the argument in NLP literature that target tasks with smaller datasets benefit the most from intermediate task fine-tuning~\cite{vu2020exploring, pruksachatkun2020intermediate, chang2021rethinking}. Since CT is the task whose training dataset has the smallest amount of samples among all four SE tasks, we believe the dataset size plays a role in explaining its boosted performance after fine-tuning other tasks (\textbf{Pattern 1}), especially CR. However, we should note that CT's performance is not always boosted by all intermediate tasks: when CR is not one of the intermediate tasks, CT's performance remains almost unchanged.

Regarding intermediate tasks, CR's dataset has the largest number of samples. Yet, it only enhances CT's performance instead of other tasks (\textbf{Pattern 1}). These findings resemble those in NLP literature, which suggest that intermediate task data size does not have consistent effect on target task performance
~\cite{pruksachatkun2020intermediate}. For \textbf{Pattern 2}, CT does not improve any other target tasks' performances. However,  Vu et al.~\cite{vu2020exploring} argued that positive transfer is still possible even when the intermediate task dataset is small. Since we do not have any other tasks whose training data is relatively the same size as CT's, it is not reasonble to draw any conclusions for this pattern. Consequently, dataset size may not adequately explain the underlying reasons for the two task ordering patterns (\textbf{Pattern 1 \& 2}) we identified.

To get a fine-grained analysis on the explanation of Pattern 1 \& 2 in terms of dataset size, we conducted an experiment where we controlled the dataset sizes of the intermediate task and target task. Similar to what Vu et al. ~\cite{vu2020exploring} did, we performed transfer experiments in four data regimes to examine the impact of dataset size on intermediate-to-target task transfer (i.e. Pattern 1 \& 2): FULL $\rightarrow$ FULL, FULL $\rightarrow$ LIMITED, LIMITED $\rightarrow$ FULL, LIMITED $\rightarrow$ LIMITED. In the FULL training regime, all training data for the associated task is used for fine-tuning. In the LIMITED regime, we randomly sampled 1K training data instances without replacement. The models that were only fine-tuned with target tasks' datasets (FULL/LIMITED) without any intermediate fine-tuning were set as baselines. Since we trained the models with 10-fold cross-validation, we conducted Welch's t-test ~\cite{ruxton2006unequal} to measure the statistical significance of the performance difference between fine-tuning chain models and baselines trained with different data regimes to identify the performance boost/drop. The experiment results for Pattern 1 and 2 are shown in Table \ref{tab:dataset-size-impact-cr-ct} and Table \ref{tab:dataset-size-impact-ct-cr} respectively. 


The LIMITED regime applied to an intermediate task shows noteworthy performance impact on the target task, even when considering its relatively small data size and short fine-tuning time in comparison to the FULL regime. We marked the score differences for each model compared to the baseline model in the Table \ref{tab:dataset-size-impact-cr-ct} and \ref{tab:dataset-size-impact-ct-cr}. For CR$\rightarrow$CT, the improvement in BLEU score compared to the CT FULL baseline is 3.361 and 1.128 for FULL$\rightarrow$FULL, LIMITED$\rightarrow$FULL respectively. In the LIMITED regime, the improvement is 25.157 and 13.385 for FULL$\rightarrow$FULL, LIMITED$\rightarrow$FULL respectively. In the case of CT$\rightarrow$CR (FULL), the drop in the score was less when CT was fine-tuned in the LIMITED regime compared to when it was fine-tuned in the FULL regime. It is important to highlight that the LIMITED regime showed a decent performance boost in CR$\rightarrow$CT and a performance drop in CT$\rightarrow$CR, even when it only made use of 9\%(1k/11.5k) from total data. This indicates that if the computing resources are insufficient or the data size is not large, the data size can be compromised to achieve a better performance.

\begin{table}[]
\centering
\caption{Dataset Size Impact Analysis Results (Pattern 1: CR→CT)}
\label{tab:dataset-size-impact-cr-ct}
\resizebox{0.9\columnwidth}{!}{%
\begin{tabular}{|lcc|lcc|}
\hline
\multicolumn{3}{|c|}{\textbf{CR→CT (FULL)}}                                                                                                 & \multicolumn{3}{c|}{\textbf{CR→CT (LIMITED)}}                                                                                              \\ \hline
\multicolumn{1}{|c|}{\textbf{Models}} & \multicolumn{1}{c|}{\textbf{Average BLEU Scores}} & \multicolumn{1}{l|}{\textbf{Score Differences}} & \multicolumn{1}{c|}{\textbf{Models}} & \multicolumn{1}{c|}{\textbf{Average BLEU Scores}} & \multicolumn{1}{l|}{\textbf{Score Differences}} \\ \hline
\multicolumn{1}{|l|}{FULL (CT)}       & \multicolumn{1}{c|}{70.881}                       & -                                               & \multicolumn{1}{l|}{LIMITED (CT)}    & \multicolumn{1}{c|}{24.841}                       & -                                               \\ \hline
\multicolumn{1}{|l|}{FULL→FULL}       & \multicolumn{1}{c|}{\textbf{74.242 (+)}}          & +3.361                                          & \multicolumn{1}{l|}{FULL→LIMITED}    & \multicolumn{1}{c|}{\textbf{49.998 (+)}}          & +25.157                                         \\ \hline
\multicolumn{1}{|l|}{LIMITED→FULL}    & \multicolumn{1}{c|}{\textbf{72.009 (+)}}          & +1.128                                          & \multicolumn{1}{l|}{LIMITED→LIMITED} & \multicolumn{1}{c|}{\textbf{38.226 (+)}}          & +13.385                                         \\ \hline
\end{tabular}%
}
\end{table}

\begin{table}[]
\centering
\caption{Dataset Size Impact Analysis Results (Pattern 2: CT→CR)}
\label{tab:dataset-size-impact-ct-cr}
\resizebox{0.9\columnwidth}{!}{%
\begin{tabular}{|lcc|lcc|}
\hline
\multicolumn{3}{|c|}{\textbf{CT→CR (FULL)}}                                                                                                 & \multicolumn{3}{c|}{\textbf{CT→CR (LIMITED)}}                                                                                              \\ \hline
\multicolumn{1}{|c|}{\textbf{Models}} & \multicolumn{1}{c|}{\textbf{Average BLEU Scores}} & \multicolumn{1}{l|}{\textbf{Score Differences}} & \multicolumn{1}{c|}{\textbf{Models}} & \multicolumn{1}{c|}{\textbf{Average BLEU Scores}} & \multicolumn{1}{l|}{\textbf{Score Differences}} \\ \hline
\multicolumn{1}{|l|}{FULL (CR)}       & \multicolumn{1}{c|}{79.566}                       & -                                               & \multicolumn{1}{l|}{LIMITED (CR)}    & \multicolumn{1}{c|}{62.965}                       & -                                               \\ \hline
\multicolumn{1}{|l|}{FULL→FULL}       & \multicolumn{1}{c|}{\textbf{76.824 (-)}}          & -2.742                                          & \multicolumn{1}{l|}{FULL→LIMITED}    & \multicolumn{1}{c|}{\textbf{77.237 (+)}}          & +14.272                                         \\ \hline
\multicolumn{1}{|l|}{LIMITED→FULL}    & \multicolumn{1}{c|}{79.336}                       & -0.23                                           & \multicolumn{1}{l|}{LIMITED→LIMITED} & \multicolumn{1}{c|}{\textbf{70.963 (+)}}          & +7.998                                          \\ \hline
\end{tabular}%
}
\end{table}

\observation{Task-specific dataset size can not sufficiently explain both positive and negative transfer task ordering patterns. Yet, the data size of fine-tuning task matters in determining the effect of a transfer. }

\subsection{Task: Affinity}
We present the results of task affinity analysis for the task orderings CR$\rightarrow$CT, and CT$\rightarrow$CR in the Table \ref{tab:task-affinity}. In addition, we include the affinity scores for the CT, and CR baseline models for comparison purposes. Note that the affinity score for a model is calculated as the sum of the loss differences between that model and the baseline model over 100 steps. 

\textit{Explanation for \textbf{Pattern 1}:}
Given the target task of CT, the intermediate fine-tuned model is the resulting model after training the pre-trained model with CR task (CR baseline in this case). The affinity score for the intermediate fine-tuned model is -6.121, while the pre-trained model has an affinity score of -6.987. The affinity score for the intermediate fine-tuned model is larger than the affinity score for the pre-trained model. This suggests that there is a low task affinity value between CR and CT because the intermediate fine-tuned model is further away from the baseline model compared to the pre-trained model. This result suggests that the training on CR has a small negative impact on the performance of the subsequent CT task. This is contradicted to the positive transfer we observed for the CR$\rightarrow$CT.


\textit{Explanation for \textbf{Pattern 2}:}
Given the target task of CR, The intermediate fine-tuned model is the resulting model after training pre-trained model with CT task (CT baseline in this case). The affinity score for the intermediate fine-tuned model is -0.763, while the pre-trained model has an affinity score of -3.471. It was observed that the affinity score for the intermediate fine-tuned model is much larger than that of the pre-trained model. This is equivalent to say that with CT, the model is further away from the baseline model compared to the pre-trained model. This result suggests that the training on CT has a large negative impact on the performance of the subsequent CR task. This matches with the negative transfer we observed for the CT$\rightarrow$CR.

\begin{table}[t]
\caption{Task Affinity Analysis Results}
\label{tab:task-affinity}
\resizebox{0.4\columnwidth}{!}{%
\begin{tabular}{|l|c|}
\hline
\textbf{Models}   & \textbf{Task Affinity Scores} \\ \hline
CR$\rightarrow$CT & -6.121                        \\ \hline
CT\_Pre-train     & -6.987                        \\ \hline
CT$\rightarrow$CR & -0.763                        \\ \hline
CR\_Pre-train     & -3.471                        \\ \hline
\end{tabular}%
}
\end{table}


\observation{Task affinity can only explain the negative transfer for CT$\rightarrow$CR, but contradicts to the positive transfer we observed for CR$\rightarrow$CT. Thus, task affinity is not a good measurement.}

\subsection{Model: Probing Tasks}

We train a logistic regression classifier for every fine-tuning chain model in this study with the probing datasets from \cite{karmakar2021pre}. Due to the space limit, we only present the probing task performances of chains that end with CT (i.e. CT is the target task) along with the fine-tuning chain model performances in Table \ref{tab:probing-ct}. The probing results for other chains are in our replication package~\cite{replication}. Recall that the mean of the classifier accuracies trained on the feature vectors from all 12 layers of the encoder was referred to as the overall probing task performance (Section \ref{sec:modelanalysis}). It is worth noting that our focus is more on the probing task performance difference between models instead of raw performance value for each model~\cite{karmakar2021pre}. Then, we conduct Spearman correlation between probing task and target task performances for every target task across different fine-tuning chain models whose task orderings end with this target task. The correlation coefficients are shown in Table \ref{tab:probingvstarget}. 

Since the performance of probing tasks can be treated as an indicator of the acquisition of particular language skills, we provide possible explanations for the task ordering patterns we found (Section \ref{sec:results}) below:

\textit{Explanation for \textbf{Pattern 2: }}The performances of fine-tuning chain models whose task orderings end with CR have a higher correlation with TYP’s performance (correlation coefficient 0.638) than those of fine-tuning chain models whose task orderings end with CT (correlation coefficient 0.342). This suggests that, for CR to perform better, the model tends to pay more attention to code semantic information when it is fine-tuned than CT does. This can also be seen from the baseline performance where TYP’s performance of CR baseline is higher than that of CT, which indicates CR learns code semantic information better than CT. Also, TYP’s performance trained on the vectors from CR baseline (90) is higher than that of fine-tuning chain model CT$\rightarrow$CR (87.833). We posit that the reason why CT$\rightarrow$CR is a negative transfer task ordering pattern is because CT doesn’t care about TYP as much as CR, so it might have hurt the subsequent CR’s effectiveness of learning code semantic information, which causes CT$\rightarrow$CR to perform worse than CR.


However, probing task performance and its correlation with target task performance can not provide enough evidence to explain why \textbf{Pattern 1} is a positive transfer pattern. It's possible that other factors such as dataset size may play a role in explaining it.

\begin{table}[t]
\centering
\caption{Correlation between Probing Task Performances and Target Task Performances}
\label{tab:probingvstarget}
\resizebox{0.7\columnwidth}{!}{%
\begin{tabular}{|l|cccc|cccl|}
\hline
\textbf{} & \multicolumn{4}{c|}{\textbf{AST}} & \multicolumn{4}{c|}{TYP} \\ \hline
\multirow{2}{*}{\begin{tabular}[c]{@{}l@{}}Correlation\\ Coefficients\end{tabular}} & CD & DD & CR & CT & CD & DD & CR & CT \\ \cline{2-9} 
 & 0.223 & 0.660 & 0.158 & 0.069 & 0.151 & 0.556 & 0.638 & 0.342 \\ \hline
\end{tabular}%
}
\end{table}

\begin{table}[ht]
\centering
\caption{Probing Task Performances and Model Performances (CT)}
\label{tab:probing-ct}
\resizebox{0.7\columnwidth}{!}{%
\begin{tabular}{|l|c|c|c|}
\hline
\textbf{\begin{tabular}[c]{@{}c@{}}Fine-tune Chain \\ Models\end{tabular}} & \textbf{\begin{tabular}[c]{@{}c@{}}AST Mean \\ Accuracy\end{tabular}} & \textbf{\begin{tabular}[c]{@{}c@{}}TYP Mean \\ Accuracy\end{tabular}} & \textbf{\begin{tabular}[c]{@{}c@{}}Model \\ Performance (BLEU)\end{tabular}} \\ \hline
CT & 77.792 & 88.500 & 70.881 \\ \hline
CR$\rightarrow$CT & 79.000 & 88.750 & 74.242 \\ \hline
CD$\rightarrow$CT & 78.958 & 87.708 & 70.879 \\ \hline
DD$\rightarrow$CT & 79.042 & 88.667 & 70.839 \\ \hline
CD$\rightarrow$CR$\rightarrow$CT & 78.875 & 88.042 & 75.058 \\ \hline
CD$\rightarrow$DD$\rightarrow$CT & 78.833 & 88.750 & 71.000 \\ \hline
CR$\rightarrow$CD$\rightarrow$CT & 77.792 & 85.708 & 73.957 \\ \hline
CR$\rightarrow$DD$\rightarrow$CT & 79.250 & 86.083 & 74.118 \\ \hline
DD$\rightarrow$CD$\rightarrow$CT & 78.708 & 87.500 & 70.739 \\ \hline
DD$\rightarrow$CR$\rightarrow$CT & 78.208 & 83.833 & 74.113 \\ \hline
CD$\rightarrow$CR$\rightarrow$DD$\rightarrow$CT & 77.083 & 86.167 & 74.204 \\ \hline
CD$\rightarrow$DD$\rightarrow$CR$\rightarrow$CT & 80.125 & 83.917 & 74.606 \\ \hline
CR$\rightarrow$CD$\rightarrow$DD$\rightarrow$CT & 79.667 & 86.750 & 73.882 \\ \hline
CR$\rightarrow$DD$\rightarrow$CD$\rightarrow$CT & 78.792 & 83.083 & 74.145 \\ \hline
DD$\rightarrow$CD$\rightarrow$CR$\rightarrow$CT & 79.792 & 83.167 & 73.950 \\ \hline
DD$\rightarrow$CR$\rightarrow$CD$\rightarrow$CT & 79.667 & 86.667 & 73.942 \\ \hline
\end{tabular}%
}
\end{table}

\observation{Intermediate tasks' learning effectiveness of code syntactic or semantic skills have an impact on the learning of the target task and its performance.}

\subsection{Model: Attention Analysis} 
\begin{figure*}[t]
    \centering
    \includegraphics[width=\textwidth]{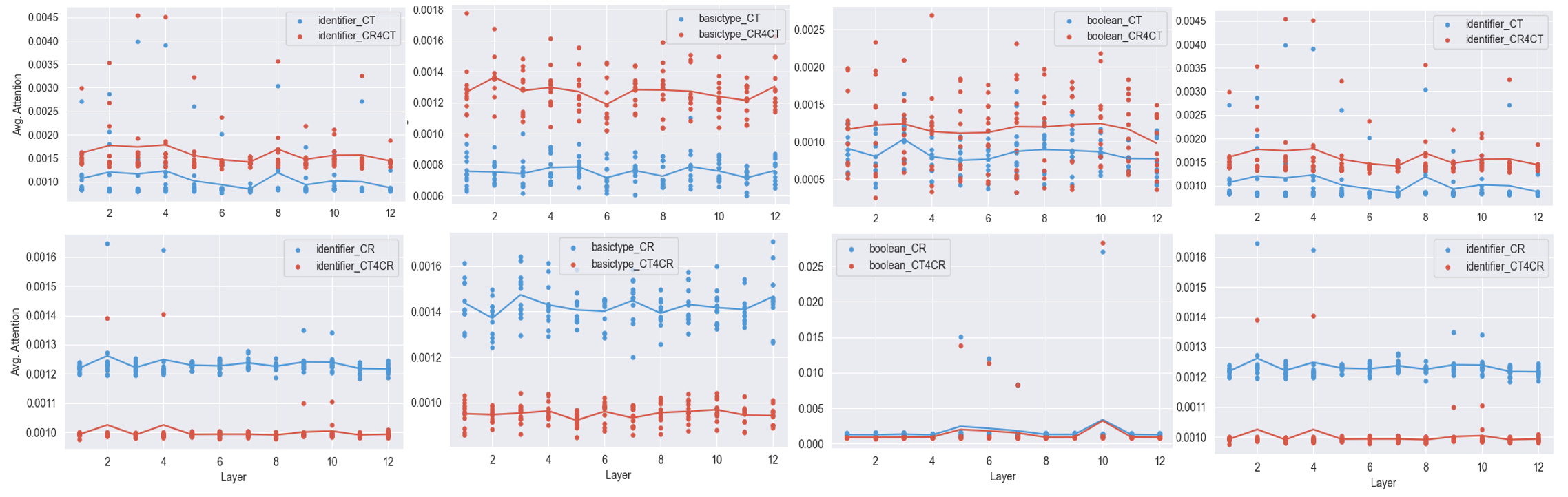}
    \caption{Attention weight experimental results on syntax tokens}
    \label{fig:attention_syntax}
\end{figure*}

\begin{figure*}[t]
    \centering
    \includegraphics[width=\textwidth]{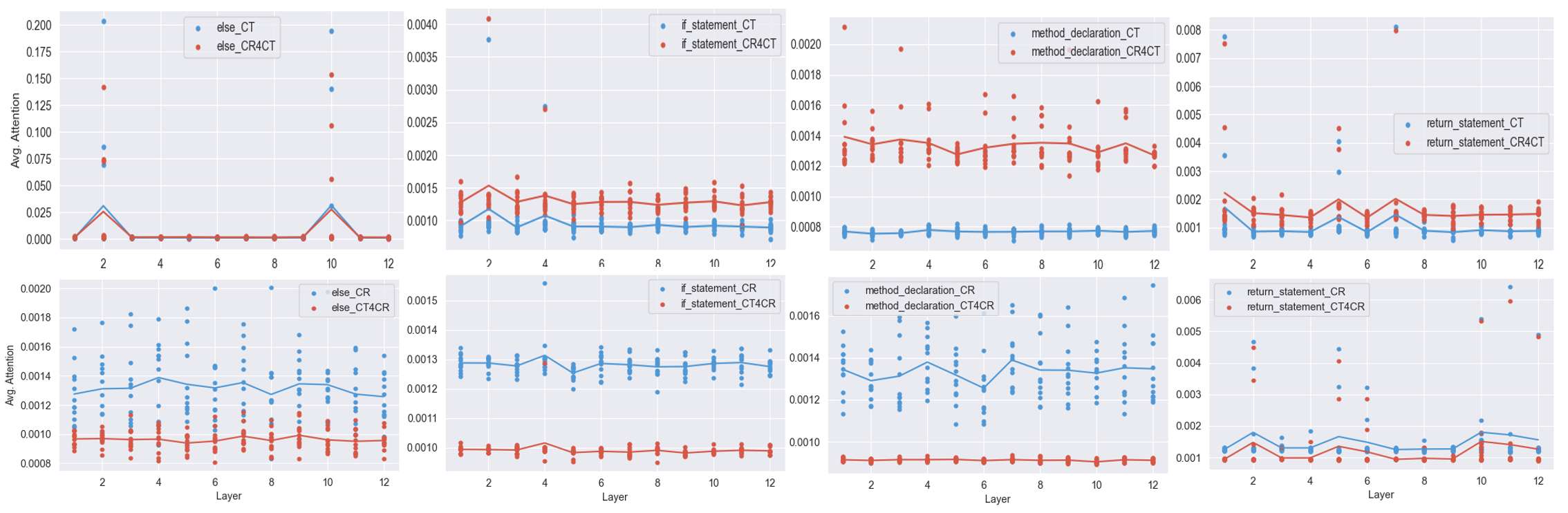}
    \caption{Attention weight experimental results on abstract syntax tree elements}
    \label{fig:attention_ast}
\end{figure*}

We present the results of our attention analysis for syntax tokens and abstract syntax tree structures in Figures \ref{fig:attention_syntax} and \ref{fig:attention_ast}, respectively. The X-axis represents the layer numbers and the Y-axis represents the attention scores. The curves in the figures indicate the average attention scores of a model across 12 layers. We observed that for the CR$\rightarrow$CT fine-tuning chain model, the average attention scores were higher than those of the CT baseline model in both syntax tokens and abstract syntax tree structure. This is consistent with the positive transfer (\textbf{Pattern 1}) we observed for CR$\rightarrow$CT task ordering. Conversely, for the CT$\rightarrow$CR fine-tuning chain model (lower sub-graphs), the average attention scores were lower than those of the CR baseline model for all cases in both syntax tokens and abstract syntax tree structure, which matches with the negative transfer (\textbf{Pattern 2}) we observed for CT$\rightarrow$CR. Based on our analysis, we can conclude that the performance of a model is positively correlated with the attention scores it receives across all 12 layers for both abstract syntax structure and syntax tokens. 



\observation{Average attention scores of the model for both abstract syntax tree structure and syntax tokens can be a factor that explains the positive and negative transfers of knowledge.}

\subsection{Time Cost Effective Analysis}
\label{sec:costeffect}
To determine the most efficient fine-tuning approaches for improving target task performance, we gathered performance gains measured by relative performance gain~\cite{vu2020exploring} (Section \ref{sec:timeanalysis}), training time (in hours), time difference between fine-tuning chain models and baseline models (Section \ref{sec:timeanalysis}), and effective ratio (Section \ref{sec:timeanalysis}) for all fine-tuning chain models that demonstrated positive transfer for target tasks. We found that the values for relative performance gain and effective ratio were too small to be easily discerned, so we multiplied them by 100 to facilitate visualization. For this analysis, we set the effective ratio threshold value at 1.697, which is the mean of all effective ratios. We consider fine-tuning chain models with an effective ratio below the threshold to be non-cost-effective. Based on our analysis, we identified seven task orderings as non-cost-effective, while the four remaining task orderings (CR$\rightarrow$CT, CD$\rightarrow$CR$\rightarrow$CT, DD$\rightarrow$CR$\rightarrow$CT, and CR$\rightarrow$DD$\rightarrow$CT) were deemed cost-effective. We recommend choosing the task orderings with effective ratios higher than the threshold value as they bring positive transfer and are cost-effective. We use mean for calculating threshold in our analysis, but it can be replaced by other standard metrics, such as median or a metric defined by the users. 


\observation{Using intermediate SE tasks may be beneficial for improving model performance, but it's crucial to consider the trade-offs between training time and performance gains. Task ordering with the maximum performance improvement may not always be the optimal choice.}

\begin{table}[t]
\caption{Performance Gain and the Corresponding Training Time}
\label{tab:time-improv}
\resizebox{0.7\columnwidth}{!}{%
\begin{tabular}{|l|c|c|c|c|}
\hline
\textbf{\begin{tabular}[c]{@{}c@{}}Fine-tune Chain \\ Models\end{tabular}} & \textbf{\begin{tabular}[c]{@{}c@{}}\textbf{Relative}\\ \textbf{Performance} \\ \textbf{Gain(\%)}\end{tabular}} & \textbf{\begin{tabular}[c]{@{}c@{}}Time\\ (Hour)\end{tabular}} & \textbf{\begin{tabular}[c]{@{}c@{}}$\Delta$Time \end{tabular}} & \textbf{\begin{tabular}[c]{@{}c@{}}Effective \\ Ratio(\%)\end{tabular}} \\ \hline
CT & / & 4 & 0 & / \\ \hline
CR$\rightarrow$CT & 4.742 & 6 & 2 & 2.371 \\ \hline
CR$\rightarrow$CD$\rightarrow$CT & 4.340 & 6.633 & 2.633 & 1.648 \\ \hline
CD$\rightarrow$CR$\rightarrow$CT & 5.893 & 6.633 & 2.633 & 2.238 \\ \hline
DD$\rightarrow$CR$\rightarrow$CT & 4.560 & 6.500 & 2.500 & 1.824 \\ \hline
CR$\rightarrow$DD$\rightarrow$CT & 4.567 & 6.500 & 2.500 & 1.827 \\ \hline
DD$\rightarrow$CR$\rightarrow$CD$\rightarrow$CT & 4.319 & 7.133 & 3.133 & 1.379 \\ \hline
DD$\rightarrow$CD$\rightarrow$CR$\rightarrow$CT & 4.330 & 7.133 & 3.133 & 1.382 \\ \hline
CR$\rightarrow$DD$\rightarrow$CD$\rightarrow$CT & 4.605 & 7.133 & 3.133 & 1.470 \\ \hline
CR$\rightarrow$CD$\rightarrow$DD$\rightarrow$CT & 4.234 & 7.133 & 3.133 & 1.351 \\ \hline
CD$\rightarrow$DD$\rightarrow$CR$\rightarrow$CT & 5.255 & 7.133 & 3.133 & 1.677 \\ \hline
CD$\rightarrow$CR$\rightarrow$DD$\rightarrow$CT & 4.688 & 7.133 & 3.133 & 1.496 \\ \hline
\end{tabular}%
}
\end{table}

\section{Implications}
\label{sec:impl}
Based on our findings, we provide the following implications for researchers and software practitioners.

\textit{\textbf{Implications for researchers:}} First, our findings in RQ1 demonstrated that some tasks' performances can either get boosted or hurt depending on the fine-tuning ordering of all tasks, while the performance may always drop for some other tasks no matter what intermediate tasks are fine-tuned before them. We believe a study on a more extensive group of SE tasks is needed to uncover more patterns and opportunities to further improve the performances of certain tasks. Second, to answer RQ2 from the perspective of task similarity, we took task affinity (Section \ref{sec:SE task sim}) as a measure in this study. However, this measure does not sufficiently explain the two task ordering patterns we identified in Section \ref{sec:results}. This might be happening due to the inability of the task affinity metric to capture task similarity. Hence, researchers should look into designing metrics to measure SE task similarity, which can efficiently reflect the transferability between SE tasks. Finally, we picked the mean effective ratio across all positive transfer fine-tuning chain models as the threshold to determine whether it's cost-effective to improve the target task performance (Section \ref{sec:costeffect}). A promising future direction would be to build a machine learning model that can automatically predict a reasonable threshold and then recommend the proper task orderings that can best improve the target task performance.

\textit{\textbf{Implications for practitioners:}} It may be intuitive to believe that the target task whose training data has a small number of data samples would always benefit from a preceding fine-tuned intermediate task with more training data, and the intermediate task with a large amount of training data may always lead to a positive transfer to downstream tasks. However, our findings suggest that this is not always the case. In order to improve task performance, we recommend practitioners to consider syntactic and semantic similarities between task training datasets when performing intermediate task fine-tuning rather than blindly trying to collect more training data. In addition, to make the model-building process more cost-efficient with intermediate task fine-tuning, practitioners can set their own effective ratio thresholds to select intermediate tasks to improve target task performance based on their resource budget.  
\section{Threats To Validity}
\label{sec:ttv}
We have taken all reasonable steps to mitigate potential threats that could hamper the validity of this study, it is still possible that our mitigation strategies might not have been effective.

\textbf{Construct validity:} We used the evaluation metrics in CodeXGLUE to assess the performance of fine-tuning chain models. If other metrics are used, the performance gain could differ. However, our goal was not to identify an absolute gain through but to highlight the possibility of performance gain, hence this threat does not impact the validity of our findings.

\textbf{Internal validity:} In this study, our probing setup does not cover all possible aspects of source code. However, to make our experiments manageable, we focused on exploring semantic and syntactic code properties learned by a model. Then, we only took two probing tasks (AST and TYP) from~\cite{karmakar2021pre} to estimate the learning effectiveness of code syntactic and semantic information by a fine-tuned model. Other variants of these two probing tasks may better estimate the model's acquisition of syntactic and semantic information. Finally, the linear classifier used for probing might be limited in its ability to learn more complex encoding of code properties. However, it was used in many probing approaches~\cite{karmakar2021pre, troshin2022probing}, demonstrating promising and reliable results.



\textbf{External validity:} Our findings may not be generalizable to all pre-trained language models of source code. Since the programming languages involved in our fine-tuning process are only Java, C/C++, and C\#, our findings may not apply to models fine-tuned with other programming languages. Moreover, we conducted our experiments on only four SE tasks, and we analyzed two task ordering patterns from them. It is possible that the conclusions from these analyses may not apply to other SE tasks, their task orderings and other models. 
\section{Conclusion and Future Work}
\label{sec:conclusion}

We conducted the first empirical study on the impact of fine-tuning intermediate SE tasks  on target task performance. We argue that the results of our fine-tuning chain model evaluations and associated discussions can provide SE researchers and practitioners with a deeper understanding of the transferability between various SE tasks, as well as the training time and performance gain trade-off when selecting task orderings for a given target task to maximize its performance.

Our study sheds light on relevant factors that may be helpful in explaining the reasons why certain task orderings (task ordering patterns) can always bring performance gain/loss to target tasks. We believe that more factors and measures should be considered to explain those patterns, and an exploration of a more extensive group of SE tasks and more pre-trained language models would be an intriguing avenue for future work. 

\section{Data Availability}
As part of our commitment to open science policy, all data collected for this study are made available as supplemental material. We provide our replication package in~\cite{replication}.
\bibliographystyle{IEEEtranS}
\bibliography{acmart}

\end{document}